# BIOPHYSICAL MODELLING OF THE EFFECTS OF INHALED RADON PROGENY ON THE BRONCHIAL EPITHELIUM FOR THE ESTIMATION OF THE RELATIONSHIPS APPLIED IN THE TWO STAGE CLONAL EXPANSION MODEL OF CARCINOGENESIS

*Balázs G. MADAS*[*], *Katalin VARGA*

*Centre for Energy Research, Hungarian Academy of Sciences, Environmental Physics Department, Konkoly-Thege Miklós út 29-33., 1121-Budapest, Hungary*

**Abstract.** There is a considerable debate between research groups applying the two stage clonal expansion model for lung cancer risk estimation, whether radon exposure affects initiation and transformation or promotion. The objective of the present study is to quantify the effects of radon progeny on these stages with biophysical models. For this purpose, numerical models of mutation induction and clonal growth were applied in order to estimate how initiation, transformation and promotion rates depend on tissue dose rate. It was found that rates of initiation and transformation increase monotonically with dose rate, while effective promotion rate decreases with time, but increases in a supralinear fashion with dose rate. Despite the uncertainty of the results due to the lack of experimental data, present study suggests that effects of radon exposure on both mutational events and clonal growth are significant, and should be considered in epidemiological analyses applying mathematical models of carcinogenesis.

INTRODUCTION

Radon is considered to be the second most important cause of lung cancer after smoking. In order to estimate cancer risk attributable to radon progeny, several epidemiological analyses were carried out. Many of them applied the two stage clonal expansion model[1], which allows the quantification of the effects of ionizing radiation on different stages of carcinogenesis, such as initiation (acquiring mutation providing a growth advantage), promotion (clonal growth of initiated cells), and transformation (acquiring mutation leading to malignancy). However, there is a considerable debate between research groups on which stages are influenced by radon exposure. Some papers argued that radon progeny acts on initiation, but not on promotion[2, 3], while others supports the hypothesis that effect on promotion plays the dominant role[4, 5]. The purpose of the present paper is to estimate the effects of radon progeny on initiation, promotion and transformation rates not by epidemiological analyses, but applying biophysical models considering initiation and transformation as mutational events, and promotion as clonal growth of mutant cells.

METHODS

Mutagenesis and clonal growth were modelled separately. The model of mutation induction is the same as presented in our earlier study[6], while the clonal expansion model is based on the work of Chao et al.[7] and Heidenreich and Paretzke[5]. The input of these models are obtained with the same cellular dosimetry model, which is similar to those applied in our earlier work[6]. A mathematical model of the bronchial epithelium was elaborated consisting of six cell types: ciliated, basal, intermediate, goblet, other secretory, and preciliated cells. Their frequency and volumes of their nuclei were obtained from experimental data[8] supposing that nuclei of preciliated cells and ciliated cells and nuclei of intermediate cells and basal cells have the same volume. Cell nuclei were represented as spheres. Only basal cells were considered as progenitors, therefore cell doses had to be computed for this cell type only. Basal cells were cylinders perpendicular to the basement membrane placed on a hexagonal lattice. The height of the cylinders is 15.0 μm, while their radius is 3.63 μm. The depth distribution in the model is based on experimental data[9], although there are significant differences, because nuclei of differentiated cells are shifted upwards, since they do not have enough place among basal cells. The experimental and model data about depth distributions of nuclei is shown in Figure 1.

---
[*] E-mail: madas.balazs@energia.mta.hu, Skype ID: madbalger

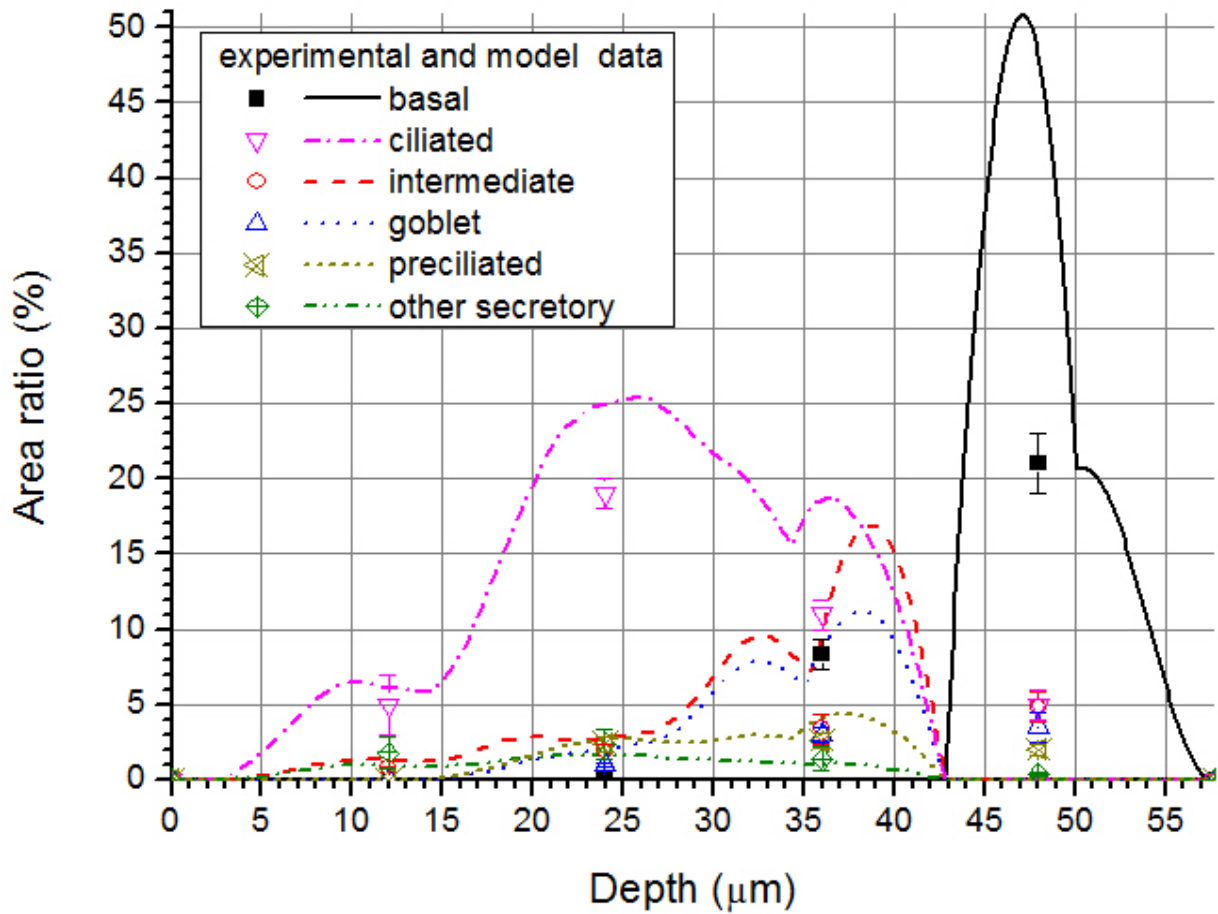

**Figure 1.** Depth-distribution of cell nuclei in the model compared to experimental data[9].

Basal cell doses and cell nucleus hits were determined by a self-developed code in an epithelium model with a size of 400 μm×400 μm×57.8 μm covered by a 5-micrometer-thick mucus layer. Radon progeny ($^{218}$Po and $^{214}$Po producing 10.4% and 89.6% of α-particles, respectively) were placed on the top of the mucus layer. β- and γ-radiation were neglected. Emission of α-particles is isotropic, and their tracks were considered as straight line segments. The deposited energy along a given segment was determined by the utilization of "The Stopping and Range of Ions in Matter" software[10]. Cell survival probability was supposed to decrease exponentially with cell nucleus hits. The slope of the curve in log-lin representation is b=-0.587 [11]. However, spontaneous death rates were also considered. In order to remain consistent with the model presented in [5], spontaneous death rate of basal cells was 0.1 year$^{-1}$.

Mutation rate was estimated by our earlier model[6] supposing that the number of induced mutations is equal to the expected number of non-repaired DNA double strand breaks (DSBs) at the next cell division of surviving cells. Both spontaneous and radiation induced DSBs were considered. The spontaneous DSB induction rate was supposed to be 0.05 hour$^{-1}$ during $G_0$ and $G_1$, and 0.10 hour$^{-1}$ during $G_2$ and M phases[12], while 50 DSB is induced during the whole S phase[13]. In the model, basal cells spend 1 hour in M, 2 hours in $G_2$, 9 hours in S phase[14], and the remaining time between subsequent cell divisions is spent in $G_0$ and $G_1$ phase. DSB-induction rate for α-particles was set to β=107 Gy$^{-1}$ [15]. Reciprocal DNA DSB repair was supposed with a characteristic time of τ=0.9 h [16]. Spontaneous death rates of differentiated cells (5.75 year$^{-1}$) were determined by the assumption that normal division rate of basal cells (1/30 day$^{-1}$)[17] maintains the equilibrium number of cells. Similarly, cell division rate ($R_{div}$) of basal cells was changed according to death rate ($R_{death}$) in order to maintain the equilibrium number of cells. This can be formulated by Equation (1):

$$R_{div} = \frac{R_{death}}{N_{pr,sd}}, \quad (1)$$

where $N_{pr,sd}$ denotes the number of progenitors survived a daily exposure. Since the adjustment of division rate to death rate is not expected immediately, the applicability of the models is limited to chronic exposures. The previous assumptions result in Equation (2) for mutation induction rate ($\dot{m}$):

$$\dot{m} = \frac{R_{div} \cdot \sum_i \exp(-b \cdot n_i) \cdot \left(0.045 \cdot \ln\left(1 + \frac{1}{\tau \cdot R_{div}}\right) + 5.99 + \beta \cdot D_i \cdot \left(0.0375d \cdot \ln\left(1 + \frac{1}{\tau \cdot R_{div}}\right) + 0.0818d\right)\right)}{\sum_i \exp(-b \cdot n_i)},$$

(2)

where $D_i$ and $n_i$ denotes the dose absorbed by the cell and the number of hits received by the nucleus of the $i^{th}$ basal cell, respectively. For each exposure rate hundred independent runs were performed.

Clonal growth was modelled by a self-developed code similar to those presented in [5] and [7]. Basal cells were placed on a hexagonal lattice. When a cell dies, its neighbours compete for the space left vacant occupying it by cell division. Since differentiated cells were not considered in this model, division means here always symmetric division. Two types of initiated cells were investigated:

Survival mutants were supposed to be partially resistant to α-radiation; their death rate was 40% of that of normal cells. This value originated from the apoptosis rate of p53-mutants of the skin exposed to UVB-radiation relative to normal cells[18]. Obviously, resistance of basal cells in the bronchi exposed to α-particles can be highly different, but more relevant data were not found in the literature. Proliferative mutants were supposed to respond more quickly for the death of their neighbours, which may correspond to some other p53-mutants having defective cell cycle checkpoints. The proliferative advantage is described by parameter g of [5]. Since the properties of proliferative mutant clones were studied in detail in [5], more attention was paid for survival mutants. Simulations started with initiation of one cell. Probability of cell death was obtained from the dosimetry model. Hundred independent runs were performed for each exposure rate simulating the first ten years after initiation. The timestep was equal to 12 hours. The average clone sizes were determined as the function of time at different exposure rates considering the extinct clones too. Effective promotion rates $\gamma_{eff}$ were computed by dividing the natural logarithm of the ratio of average clone sizes ($N(t_2)$ and $N(t_1)$) at $t_2$ and $t_1$ by $t_2-t_1$:

$$\gamma_{eff} = \frac{\ln\left(\frac{N(t_2)}{N(t_1)}\right)}{t_2 - t_1} \quad . \quad (3)$$

RESULTS AND DISCUSSION

In Figure 2, mechanisms of mutation induction by radiation are compared. The effects of accelerated turnover and DNA damages were excluded mathematically by writing the background division rate instead of $R_{div}$ and zeros instead of $D_i$-s, respectively. Mutation rate is dominated by the accelerated turnover of basal cells due to the death of other cells. In the range shown, mutation rate depends linearly on dose rate. In the present scenario, most cells killed are differentiated, because the mean depth of progenitors is higher as compared to our earlier work [6]. In addition, higher mutation rates were obtained than earlier, because the relative number of progenitors was lower, and higher values of the available experimental data were chosen both for b and β. Several epidemiological studies on lung cancer among uranium miners supposed that cell killing effect of radiation decreases initiation and transformation rates applying such functions for these steps, which include a factor decreasing exponentially by exposure rate [19, 20]. Any decrease due to cell death in initiation and transformation rates with exposure rate is not supported by our model showing monotonic increase of the function between mutation rate and dose rate. This is in agreement with one conclusion of the cited studies [19, 20] that there is no indication of a statistically significant cell-killing effect on the first mutation.

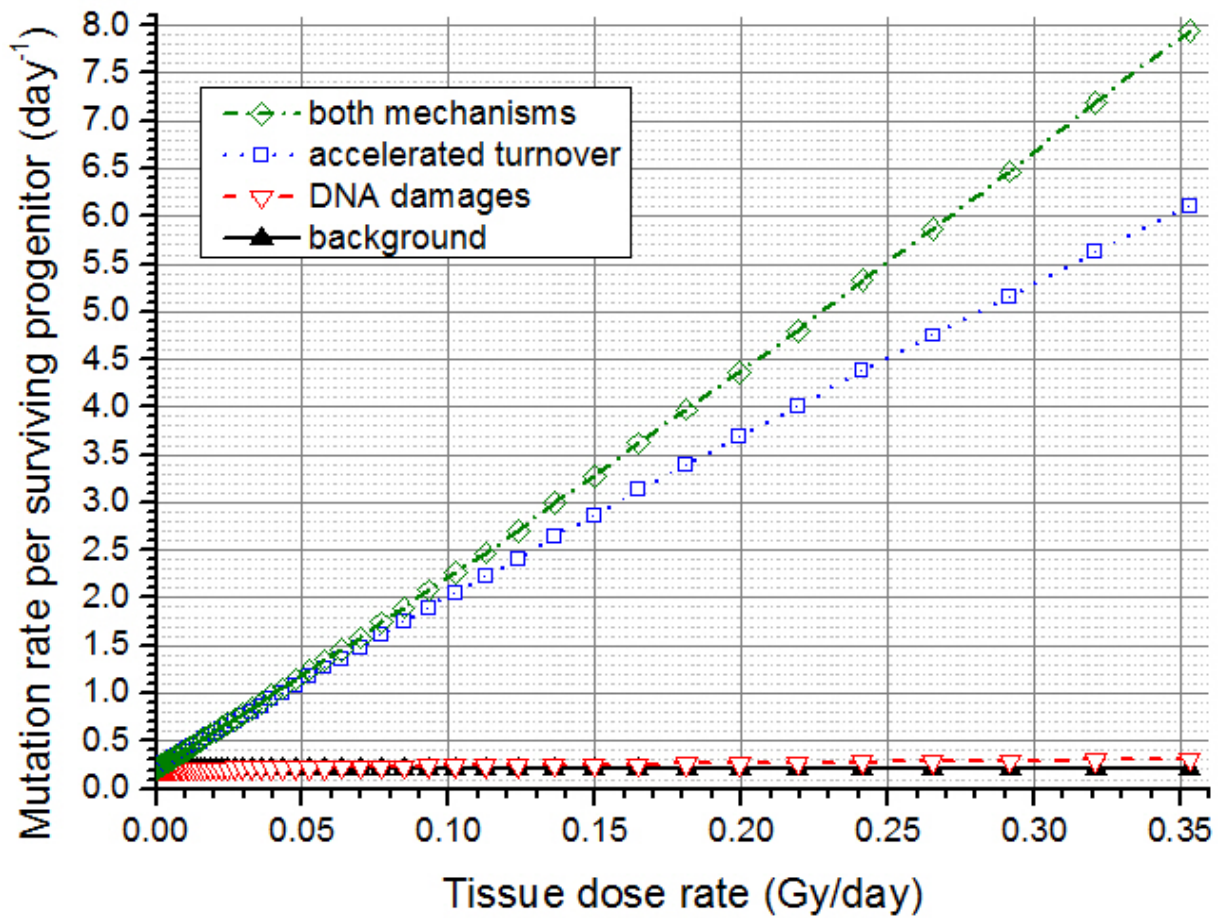

**Figure 2.** Mutation induction rate as the function of tissue dose rate and the contribution of accelerated cell turnover of progenitors and radiation induced DNA damages in progenitors. Spontaneous mutation rate obtained by the model is also plotted.

In Figure 3, average clone sizes presented by the number of mutant cells are shown as the function of tissue dose rate ten years after initiation (upper panel) and as the function of time since initiation at different tissue dose rates (bottom panel). Number of cells in a clone increases quadratically with dose rate and time, and so with cumulative dose too. This is because proliferation is limited to the periphery of the clone, and its perimeter increases linearly with dose rate and time. This growth is much slower than exponential, which means that effective promotion rate decreases with clone size. However, effective promotion rate increases, but its slope decreases with dose rate similarly to the functions applied in epidemiological studies [21, 22]. Proliferative mutants characterised by several values for *g* were tested in order to find that, which provides growth properties similar to survival mutants having 40% of death rate of normal cells. Nice agreement was found at *g*=0.12, as it is shown in Figure 3. However, since reproductive clones extinct more frequently than surviving clones, the average size of living reproductive clones is much higher than the average size of living surviving clones.

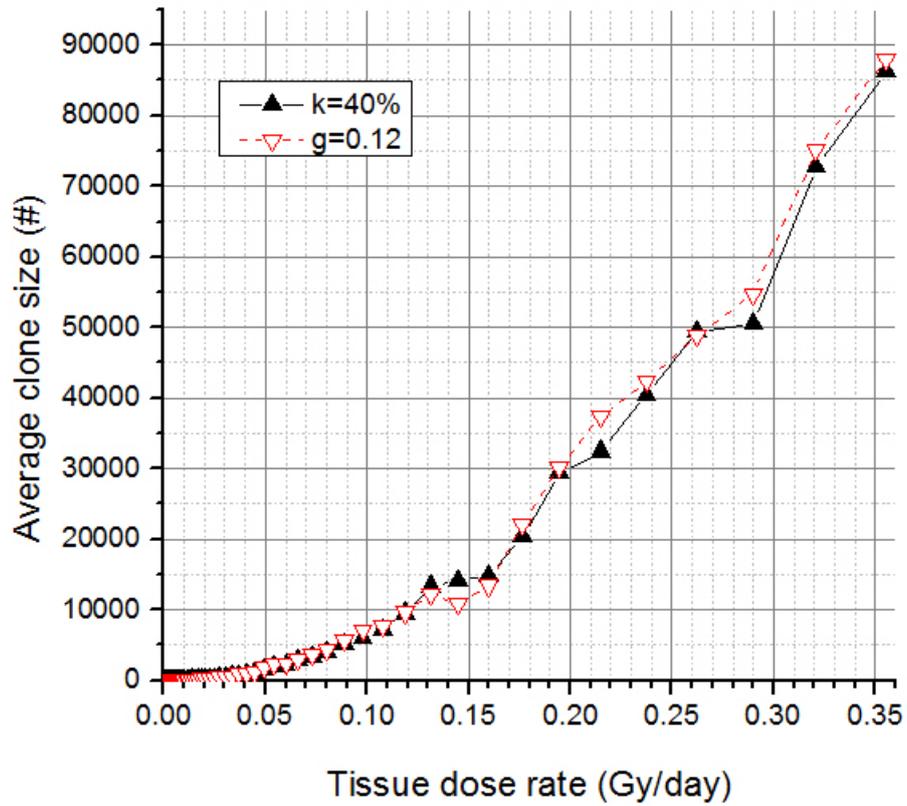
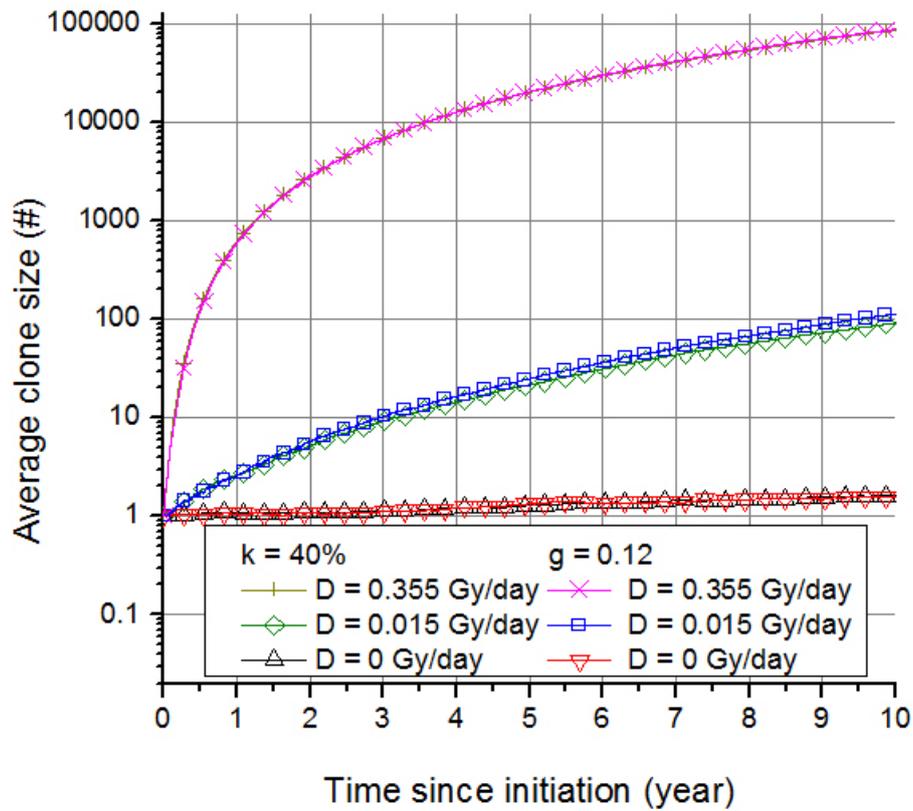

**Figure 3.** Average clone size 10 years after initiation as the function of tissue dose rate (upper panel) and average clone size as the function of time since initiation at different dose rates (bottom panel). Average clone size including extinct clones is present.

Figure 4 shows the relative increase in initiation and transformation rates and in effective promotion rate due to radon exposure as the function of tissue dose rate. Initiation and transformation may require several mutations, which were considered by plotting the square and the cube of mutation induction rate too. Since promotion rate is not constant along time, effective promotion rate is computed for different time scales. If initiation corresponds to one mutation, then the significance of promotion is much higher in the present scenario. However, if several mutations were necessary for initiation (and/or transformation), then they would dominate the process even at relatively low cumulative doses. Based on the present study, both mutational events and clonal growth is affected by radon exposure and should not be neglected in epidemiological analyses applying the two stage clonal expansion model.

Since the deposition of radon progeny is highly heterogeneous in the lungs, and especially in the bronchi [23], it remains an important question whether bronchial carcinoma originates from the deposition hot spots or from the rest of the bronchi. Importantly, 0.35 Gy/day tissue dose rate at the most exposed parts of the bronchi corresponds to approximately 0.7 WL if eight hours per day is spent in this exposure rate [6]. Studies on the effect of the gradient in dose rate, and development of methods for consideration of spatial dose distribution in epidemiological studies may be necessary for better estimation of risk related to inhalation of radon progeny.

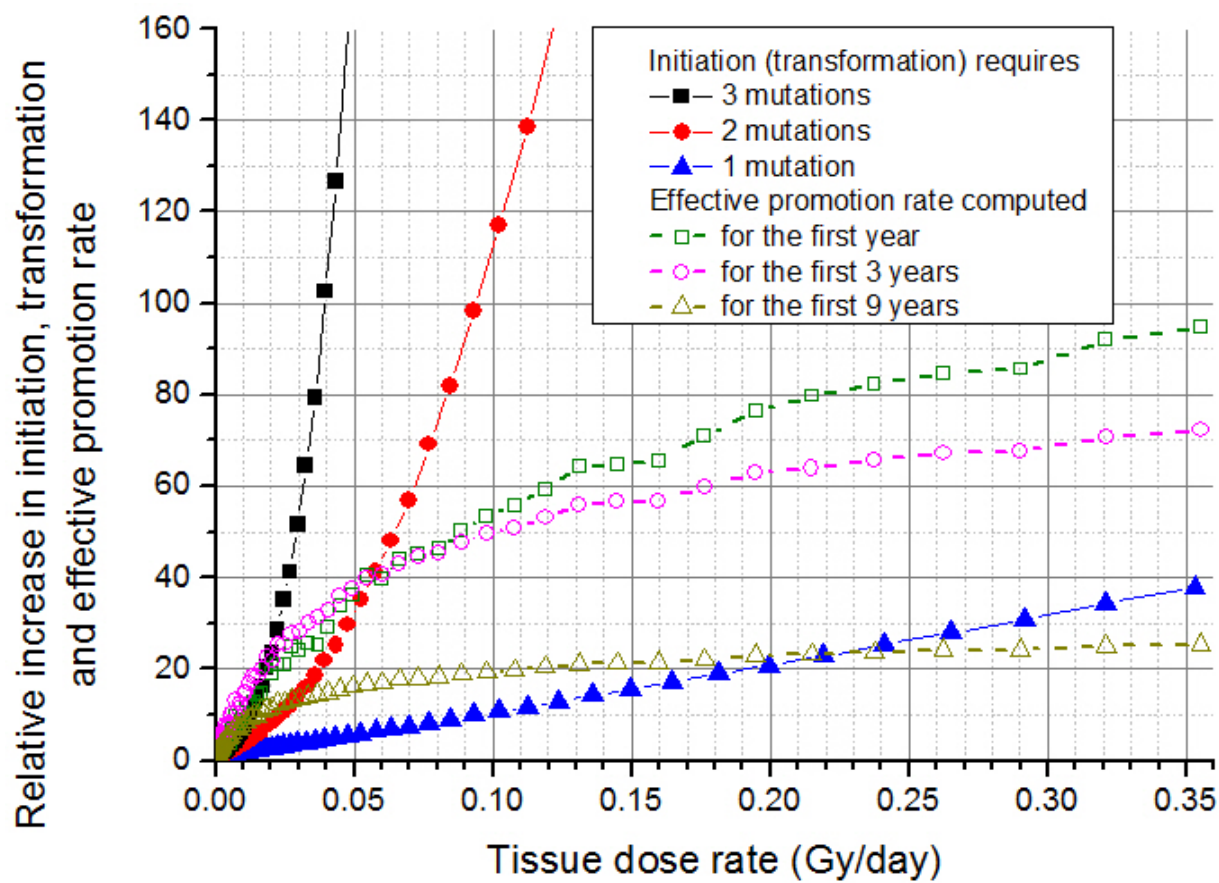

**Figure 4.** Relative increase in rates of initiation and transformation if they require one, two, or three mutations and the relative increase in effective promotion rate computed for the first, the first three, and the first nine years after initiation as the function of tissue dose rate.

CONCLUSIONS

Epidemiological analyses on lung cancer among uranium miners applying the two stage clonal expansion model present inconsistent findings about the effects of radon exposure on initiation, promotion, and transformation. In the present study, we applied biophysical models of mutation induction and clonal growth in order to estimate the dose rate dependence of initiation, transformation and promotion. Rates of initiation and transformation being considered as mutational events increase monotonically with dose rate. Cell death induced by radiation enhances not only the promotion rate, but initiation and transformation rates as well challenging the assumption of several epidemiological analyses. Based on the present study, polynomial functions should be applied in mathematical modelling of carcinogenesis in uranium miners for describing the dependence of initiation and transformation on exposure rate of radon progeny. The degree of the polynomial depends on the number of mutations required for initiation and transformation. Since the growth of the clone is limited to its periphery, clone size increases quadratically. Effective promotion rate and effective promotion rate relative to effective promotion rate in unexposed tissue decreases with observation time, but increases in a supralinear fashion with dose rate similarly to the outcome of several epidemiological analyses. Despite the uncertainty of the results due to the lack of experimental data on input parameters, the present study suggest that exposure to radon progeny affects significantly both mutational events and growth of mutant clones in the lungs, and these effects should not be neglected in epidemiological analyses applying mathematical models of carcinogenesis. Since the exposure of bronchi to radon progeny is spatially very inhomogeneous, further studies are required to determine whether lung cancer originates from small parts of the epithelium characterised by high dose rates or from large parts of the epithelium exposed to low dose rates. The effects of the gradient in dose rate on clonal growth should be also investigated.

REFERENCES

1. Armitage, P. and Doll, R. A two-stage theory of carcinogenesis in relation to the age distribution of human cancer. British Journal of Cancer. **11**(2), 161-169 (1957).
2. Bijwaard, H., Brugmans, M. J. P. and Schollnberger, H. Can promotion of initiated cells be explained by excess replacement of radiation-inactivated neighbor cells? Radiation Research. **165**(6), 741-744 (2006).
3. Brugmans, M. J. P., Bijwaard, H. and Leenhouts, H. P. The overrated role of 'promotion' in mechanistic modelling of radiation carcinogenesis. Journal of Radiological Protection. **22**(3), A75-79 (2002).
4. Heidenreich, W. F., Atkinson, M. and Paretzke, H. G. Radiation-induced cell inactivation can increase the cancer risk. Radiation research. **155**(6), 870-872 (2001).
5. Heidenreich, W. E. and Paretzke, H. G. Promotion of Initiated Cells by Radiation-Induced Cell Inactivation. Radiation Research. **170**(5), 613-617 (2008).
6. Madas, B. G. and Balásházy, I. Mutation induction by inhaled radon progeny modeled at the tissue level. Radiation and Environmental Biophysics. **50**(4), 553-570 (2011).
7. Chao, D. L., Eck, J. T., Brash, D. E., Maley, C. C. and Luebeck, E. G. Preneoplastic lesion growth driven by the death of adjacent normal stem cells. Proceedings of the National Academy of Sciences of the United States of America. **105**(39), 15034-15039 (2008).
8. Mercer, R. R., Russell, M. L., Roggli, V. L. and Crapo, J. D. Cell number and distribution in human and rat airways. American Journal of Respiratory Cell and Molecular Biology. **10**(6), 613-624 (1994).
9. Mercer, R. R., Russell, M. L. and Crapo, J. D. Radon dosimetry based on the depth distribution of nuclei in human and rat lungs. Health Physics. **61**(1), 117-130 (1991).
10. Ziegler, J. F., Biersack, J. P. and Ziegler, M. D. SRIM - The Stopping and Range of Ions in Matter. Ion Technology Press (2008).
11. Soyland, C. and Hassfjell, S. P. Survival of human lung epithelial cells following in vitro alpha-particle irradiation with absolute determination of the number of alpha-particle traversals of individual cells. International Journal of Radiation Biology. **76**(10), 1315-1322 (2000).
12. Hazelton, W. D. Modeling the effects of radiation on cell cycle regulation and carcinogenesis. (Singapore: World Scientific) (2008).
13. Vilenchik, M. M. and Knudson, A. G. Endogenous DNA double-strand breaks: Production, fidelity of repair, and induction of cancer. Proceedings of the National Academy of Sciences of the United States of America. **100**(22), 12871-12876 (2003).
14. Ijiri, K. and Potten, C. S. The circadian-rhythm for the number and sensitivity of radiation-induced apoptosis in the crypts of mouse small-intestine. International Journal of Radiation Biology. **58**(1), 165-175 (1990).
15. Claesson, A. K., Stenerlow, B., Jacobsson, L. and Elmroth, K. Relative biological effectiveness of the alpha-particle emitter At-211 for double-strand break induction in human fibroblasts. Radiation Research. **167**(3), 312-318 (2007).
16. Fowler, J. F. Repair between dose fractions: A simpler method of analyzing and reporting apparently biexponential repair. Radiation Research. **158**(2), 141-151 (2002).
17. Adamson, I. Y. R. Cellular kinetics of the lung. (Berlin: Springer) (1985).
18. Ziegler, A., Jonason, A. S., Leffell, D. J., Simon, J. A., Sharma, H. W., Kimmelman, J., Remington, L., Jacks, T. and Brash, D. E. Sunburn and p53 in the onset of skin-cancer. Nature. **372**(6508), 773-776 (1994).
19. Brugmans, M. J. P., Rispens, S. M., Bijwaard, H., Laurier, D., Rogel, A., Tomasek, L. and Tirmarche, M. Radon-induced lung cancer in French and Czech miner cohorts described with a two-mutation cancer model. Radiation and Environmental Biophysics. **43**(3), 153-163 (2004).
20. van Dillen, T., Dekkers, F., Bijwaard, H., Kreuzer, M. and Grosche, B. Lung Cancer from Radon: A Two-Stage Model Analysis of the WISMUT Cohort, 1955-1998. Radiation Research. **175**(1), 119-130 (2011).
21. Eidemuller, M., Jacob, P., Lane, R. S. D., Frost, S. E. and Zablotska, L. B. Lung Cancer Mortality (1950-1999) among Eldorado Uranium Workers: A Comparison of Models of Carcinogenesis and Empirical Excess Risk Models. Plos One. **7**(8)(2012).
22. Heidenreich, W. F., Tomasek, L., Grosche, B., Leuraud, K. and Laurier, D. Lung cancer mortality in the European uranium miners cohorts analyzed with a biologically based model taking into account radon measurement error. Radiation and Environmental Biophysics. **51**(3), 263-275 (2012).
23. Balásházy, I., Farkas, Á., Madas, B. G. and Hofmann, W. Non-linear relationship of cell hit and transformation probabilities in a low dose of inhaled radon progenies. Journal of Radiological Protection. **29**(2), 147-162 (2009).